\documentclass[11pt, a4paper]{article}
\usepackage{jcappub}

\usepackage[utf8]{inputenc} 
\usepackage[T1]{fontenc}    
\usepackage{lmodern}
\usepackage{hyperref}       
\usepackage{url}            
\usepackage{amsfonts}       
\usepackage{amsmath,amsthm,amssymb} 
\usepackage{bm}
\usepackage{bbm}
\usepackage{dcolumn}
\usepackage{nicefrac}       
\usepackage{graphicx}
\usepackage{xcolor, etoolbox}
\usepackage{mathtools}
\usepackage{mathrsfs}
\usepackage{soul}
\usepackage{ragged2e}
\usepackage{microtype}      
\usepackage{braket}      
\usepackage{enumitem}
\usepackage{xspace}
\usepackage{dsfont}
\usepackage{float}
\usepackage[export]{adjustbox}

\usepackage[normalem]{ulem} 

\DeclarePairedDelimiter\abs{\lvert}{\rvert}
\renewcommand{\vec}[1]{\boldsymbol{#1}}

\hypersetup{
    colorlinks,
    linkcolor=blue,
    citecolor=blue,
    urlcolor=blue,
    breaklinks=true 
}

\graphicspath{{./figures/}} 

\begin{document}

\title{\boldmath Vector dark matter production during inflation and reheating}

\author[a]{Jose A. R. Cembranos,}
\author[a]{Luis J. Garay,}
\author[a]{Álvaro Parra-López}
\author[b]{and Jose M. Sánchez Velázquez}

\affiliation[a]{Departamento de F\'isica Te\'orica and IPARCOS, Facultad de Ciencias F\'isicas, \\
Universidad Complutense de Madrid, Ciudad Universitaria, 28040 Madrid, Spain} 
\affiliation[b]{Instituto de F\'isica Te\'orica UAM/CSIC, c/ Nicol\'as Cabrera 13-15,\\ Cantoblanco, 28049, Madrid, Spain}

\emailAdd{cembra@fis.ucm.es}
\emailAdd{luisj.garay@ucm.es}
\emailAdd{alvaparr@ucm.es}
\emailAdd{jm.sanchez.velazquez@csic.es}
	
\date{\today}

\abstract{Gravitational particle production of spectator fields due to the expansion universe during the inflationary and reheating phases of the early universe is of particular interest in the context of dark matter, since it allows to constrain the properties of the dark candidate by comparing the density of particles produced with the observed dark matter abundance. In such processes, tachyonic instabilities arise as a consequence of the coupling to the curvature, greatly enhancing mode production. In this work, we consider a massive vector field that is coupled to the curvature scalar and the Ricci tensor only, and study its gravitational production through inflation and reheating. We show how the mechanism is more efficient than in the case of a non-minimally coupled scalar field, giving rise to larger abundances. Moreover, we analyze the importance of the coupling to the Ricci tensor, which increases tachyonic instabilities in the system, and constrain the mass of the dark particle and the values of the coupling constants by comparing the corresponding abundance with observations.}


\arxivnumber{-}
~\hfill IPARCOS-UCM-23-119\par
~\hfill IFT-UAM/CSIC-23-128\par
\maketitle
\flushbottom


\section{Introduction}
\label{sec:introduction}  

Although dark matter is one of the keystones of our modern explanation of astrophysical and cosmological phenomena, there are still many fundamental properties about its own nature that are unknown. Observational evidence suggests that any suitable dark matter model must interact weakly with the Standard Model fields, it must be non-relativistic in the present, and furthermore, it must have appropriate clustering properties in order to explain large scale structure formation  \cite{Sofue2001,Bartelmann2001,Clowe2004,Planck2018}. However, due to the rather indirect nature of the observational evidence, fundamental properties such as the mass or the spin of the dark matter field are unknown and more theoretical work must be done in order to unveil them.

Within the current scenario, it becomes imperative to explore different paths that may shed light into its fundamental properties. One of the most paved ways in the last years has been to explore how different creation mechanism for dark matter particles can constrain the hypothetical space of parameters for its mass, its coupling to gravity and other fields, or even its spin. Within all the theoretical proposals for plausible creation mechanisms, we want to draw attention to one which is inherent to quantum fields that have only gravitational interactions: gravitational particle production. Due to the dynamical nature of space-time, any field will undergo particle creation \cite{Parker1969,Ford1987,Ford2021}. Moreover, this phenomenon becomes particularly important during the stages of inflation \cite{Starobinsky1980, Guth1981, Linde1982} and reheating \cite{Kolb1990,Kofman1994,Kofman1997,Allahverdi2010,Baumann2015} in the early universe, since the geometry of spacetime rapidly changes. Furthermore, if the field under consideration is free, there is no mechanism to dilute the created particles, as no interaction with other fields is allowed, hence making it one of the most appealing mechanisms to explore for dark matter candidates. 

These ideas have been applied mainly to scalar fields \cite{Chung1998,Chung2001,Markkanen2018,Fairbairn2019,Chung2019,Hashiba2019,Herring2020,Kainulainen2023,Garcia2023}, but also to vectors and fermions in the references \cite{Graham2015,Ema2019,Bastero2019,Ahmed2020,Sato2022}. Moreover, recent works have shown that if one wants to consider gravitational production to obtain robust constraints on the parameters of the field, a detailed characterization of spacetime dynamics is needed. Indeed, it has been analyzed in the few past years how the long-time ignored oscillations that the curvature undergoes during reheating or the specific slow-roll dynamics during inflation can change the theoretical production of particles by several orders of magnitude \cite{Ema2016,Markkanen2017a,Ema2018,Cembranos2020,ScalarField2023}. Hence it is natural to revisit the production of vector fields as dark matter candidates, using a more sophisticated description of the background geometry during the inflationary and reheating epochs of our universe.

In this work, we analyze the gravitational production of a vector dark matter field which is only coupled to gravity (through the curvature scalar and the Ricci tensor) in the most generic way considering only renormalizable terms. The inflationary model considered involves a single inflaton field $\phi$ that slowly rolls down a quadratic potential and oscillates around its minimum in the transition to reheating. The dynamics is obtained after numerically solving the corresponding equation of motion without any approximations. Note that this model, in its simplest version, is ruled out by CMB observations, although it allows for comparison with previous literature, in particular with the recent work \cite{ScalarField2023}. Nevertheless, the procedure can be applied similarly to other choices of potential. We consider that the spectator dark matter field is initially prepared in the Bunch-Davies state, and particle production takes place during the expansion of the geometry. This will generate a non-vanishing, comoving particle density that will reach a constant value once the expansion of the geometry becomes adiabatic. In order to extract this quantity, we solve the mode equation until the adiabatic regime is reached. Then we use the customary adiabatic vacuum prescription for obtaining the Bogoliubov coefficients relating the latter with the \textit{in} vacuum. Additionally, we use the analytical slow-roll approximation to the mode equation derived in \cite{ScalarField2023} in order to speed up the numerical calculations. Within our derivation, we can set constraints on the possible values for each of the coupling strengths and the mass of the produced quanta so that it does not lead to overproduction, by comparing the resulting abundance with observations.

The remainder of this paper is organized as follows. In section \ref{sec:vectorfield}, we describe the dynamics of a massive vector field in an expanding cosmology, and write its mode content in terms of an effective mass tensor that incorporates all couplings to the geometry. At the same time, we obtain the mode equations and characterize the dynamics of the inflaton. Particle production of transverse and longitudinal modes is obtained in sections \ref{sec:particleproductionT} and \ref{sec:particleproductionL}, respectively, whereas we perform an analysis of the total abundance yield in section \ref{sec:abundances}. Lastly, we elaborate our conclusions in section \ref{sec:conclusions}.

\emph{Notation.} We set $M_{\text{P}}=1/\sqrt{G}, \hbar = c = k_{\text{B}} = 1$, and use the metric signature $\left(-, +, +, +\right)$. Furthermore, greek indices $\mu, \nu$ run from $0$ to $3$, while latin indices $i,j$ run from $1$ to $3$.

\section{Dynamics of a massive vector field in flat FLRW cosmologies}
\label{sec:vectorfield}

Let us consider a massive vector field $A_{\mu}$ in a flat Friedmann-Lemaître-Robertson-Walker (FLRW) geometry \cite{Friedman1922,Friedman1924,Lemaitre1931,Robertson1935,Robertson1936a,Robertson1936b,Wald1984}  which is non-minimally coupled to the geometry, extending therefore the minimally coupled model presented in references~\cite{Ema2019, Bastero2019, Ahmed2020}. 

The interaction with gravity is taken into account through terms of the form $RA_{\mu}A^{\mu}$ and $\Tilde{R}^{\mu\nu}A_{\mu}A_{\nu}$ in the action
\begin{equation}
S = -\frac{1}{2} \int \text{d}^4x \sqrt{-g} \left(\frac{1}{2}F_{\mu\nu}F^{\mu\nu} + m^2 A_{\mu}A^{\mu} + \gamma R A_{\mu}A^{\mu} + \sigma \Tilde{R}^{\mu\nu} A_{\mu}A_{\nu}\right),
\label{eq:GeneralAction}
\end{equation}
where $g$ is the determinant of the metric, $m$ is the mass of the boson, $F_{\mu\nu}=\partial_{\mu}A_{\nu} - \partial_{\nu}A_{\mu}$ is the field strength, and $\gamma$ and $\sigma$ are the couplings to the Ricci scalar $R$ and the traceless Ricci tensor $\Tilde{R}^{\mu\nu} = R^{\mu\nu} - g^{\mu\nu}R/4$, respectively. The FLRW metric can be written, considering conformal time $\eta$ and using Cartesian coordinates for the spatial sections, in a conformally flat form,
\begin{equation}
g_{\mu\nu} = a^2(\eta) \eta_{\mu\nu},
\label{eq:Metric}
\end{equation}
where cosmological time $t$ follows from $dt=a(\eta)d\eta$, and $\eta_{\mu\nu}$ is the Minkowski metric, which will be used in the following for raising and lowering indices. Introducing \eqref{eq:Metric} in \eqref{eq:GeneralAction}, we can explicitly write
\begin{equation}
S = -\frac{1}{2}\int d^4x \left[\frac{1}{2}F^{\mu\nu}F_{\mu\nu} + M^{\mu\nu}A_{\mu}A_{\nu}\right],
\label{eq:ExplicitAction}
\end{equation}
where we have defined the mass tensor $M^{\mu\nu}$ as
\begin{equation}
M^{\mu\nu} \equiv a^2\eta^{\mu\nu}\left(m^2 + \gamma R\right) + \sigma\Tilde{R}^{\mu\nu}.
\end{equation}
This action leads to the equation of motion
\begin{equation}
\partial_{\mu}\partial^{\mu}A^{\nu} - \partial_{\mu}\partial^{\nu}A^{\mu} =M^{\mu\nu}A_{\mu},
\label{eq:ProcaEquation}
\end{equation}
which is a generalization of the \textit{Proca equation} for massive bosons \cite{Peskin1995} (the latter being recovered in the case of vanishing couplings $\gamma$ and $\sigma$).

Homogeneity allows us to we write the field $A_{\mu}$ in momentum space through Fourier modes,
\begin{equation}
A_{\mu}(\eta, \vec{x}) = \int \frac{d^3k}{(2\pi)^{3/2}} \Tilde{A}_{\mu}(\eta, \vec{k})e^{i\vec{k}\cdot\vec{x}},
\end{equation}
with the conjugate field fulfilling $\Tilde{A}_{\mu}(\eta, \vec{k}) = \Tilde{A}^*_{\mu}(\eta, -\vec{k})$ in order to ensure that $A_{\mu}$ is real. From now on, we will work with $\Tilde{A}_{\mu}$ but drop the tilde for clarity. Additionally, spatial components are divided in \textit{longitudinal} and \textit{transverse} modes defined as
\begin{equation}
\vec{A} = \vec{A}_{\text{T}} + A_\text{L} \frac{\vec{k}}{\abs{k}},
\end{equation}
with 
\begin{equation}
\vec{k} \cdot \vec{A} = k A_{\text{L}} \qquad \text{and} \qquad \vec{k} \cdot  \vec{A}_{\text{T}} = 0.
\end{equation}
Writing~\eqref{eq:ProcaEquation} in momentum space, we observe that $A_0$ is not dynamical, but can actually be expressed in terms of the longitudinal part,
\begin{equation}
\left[k^2 - M^{00}\right]A_0(\eta, \vec{k}) = -ikA_{\text{L}}^{\prime}(\eta, \vec{k}).
\end{equation}
Here, the prime denotes derivative with respect to conformal time. Moreover, as long as~$k^2 - M^{00}$ is different than $0$, one can write
\begin{equation}
A_0(\eta, \vec{k}) = -\frac{ikA_{\text{L}}^{\prime}(\eta, \vec{k})}{k^2 - M^{00}}.
\label{eq:0Component}
\end{equation}
This condition requires that $M^{00}$ be negative for all values of $m, \gamma$, and $\sigma$, which restricts the allowed parameter space of our theory. Note that this is indeed a fundamental limitation of the theory, and not an operational requirement, since positive values of $M^{00}$ would lead to instability problems (see equation \eqref{eq:MomentumAction} below). Since the explicit value of $M^{00}$ depends on the background dynamics, we will analyze the allowed region of parameters in subsection~\ref{subsec:background}.  

Expanding \eqref{eq:ExplicitAction} in Fourier modes and making use of \eqref{eq:0Component}, we arrive at
\begin{equation}
S = \frac{1}{2}\int \frac{d\eta \, d^3k}{(2\pi)^{3/2}}\Bigg[\abs{\vec{A}_{\text{T}}^{\prime}}^2- \left(k^2 + M^{\text{TT}}\right)\abs{\vec{A}_{\text{T}}}^2-\frac{M^{00}}{k^2 -M^{00}}\abs{\vec{A}_{\text{L}}^{\prime}}^2 -M^{\text{LL}}\abs{\vec{A}_{\text{L}}}^2\Bigg],
\label{eq:MomentumAction}
\end{equation}
We note that the dynamics of transverse and longitudinal degrees of freedom are independent of each other, and as such can be treated separately. Moreover, the fact that a positive $M^{00}$ leads to instabilities is clear in \eqref{eq:MomentumAction}, since in this case, the kinectic term of the longitudinal mode becomes negative.

\subsection{Dynamics of transverse modes}

 The action for the transverse modes is given by
\begin{equation}
S_{\text{T}} = \frac{1}{2}\int \frac{d\eta \, d^3k}{(2\pi)^{3/2}}\left[\abs{\vec{A}_{\text{T}}^{\prime}}^2 - \left(k^2 + M^{\text{TT}}\right)\abs{\vec{A}_{\text{T}}}^2\right],
\label{eq:TransverseAction}
\end{equation}
and leads to the equation of motion
\begin{equation}
\vec{A}_{\text{T}}^{\prime\prime}(\eta, \vec{k}) + \left[k^2 + M^{\text{TT}}(\eta)\right]\vec{A}_{\text{T}}(\eta, \vec{k}) = 0.
\end{equation}
The most general solution can be written as
\begin{equation}
\vec{A}_{\text{T}}(\eta, \vec{k}) = \sum_{h=\pm}\left[a_{\text{T}, h}(\vec{k}) v_{\text{T}}(\eta, k)\vec{\epsilon}_h  +  a^*_{\text{T}, h}(-\vec{k})v_{\text{T}}^*(\eta, k) \vec{\epsilon}^*_h\right],
\label{eq:TransverseExpansion}
\end{equation}
where $\vec{\epsilon}_h$ is a polarization vector ($h=+1$ or $-1$). Note that this expression fulfills the condition $\vec{A}^*_{\text{T}}(\eta, \vec{k}) = \vec{A}_{\text{T}}(\eta, -\vec{k})$, as required by the reality of $A_{\mu}$. In turn, this implies that~$v_{\text{T}}(\eta, k)$ only depends on the norm of the $3$-momentum. Therefore, the mode equation for $\vec{A}_{\text{T}}$ boils down to
\begin{equation}
v_{\text{T}}^{\prime\prime}(\eta, k) + \omega_{\text{T}}^2(\eta, k)v_{\text{T}}(\eta, k) = 0,
\label{eq:TransverseModeEquation}
\end{equation}
where
\begin{equation}
   \omega_{\text{T}}^2(\eta, k)=k^2 + M^{\text{TT}}(\eta)
\label{eq:TransverseFrequency}
\end{equation}
and
\begin{equation}
    M^{\text{TT}} = a^2\left(m^2 + \gamma R\right) + \sigma\Tilde{R}^{\text{TT}}.
\end{equation}
This has the shape of a harmonic oscillator equation with a time-dependent frequency, and for $\sigma=0$, it has exactly the shape of the equation of the non-minimally coupled scalar field discussed in~\cite{ScalarField2023}.

\subsection{Dynamics of longitudinal modes}

On the other hand, the action for the longitudinal mode reads
\begin{equation}
S_{\text{L}} = \frac{1}{2}\int \frac{d\eta d^3k}{(2\pi)^{3/2}}\left[-\frac{M^{00}}{\vec{k}^2 - M^{00}}\abs{\vec{A}_{\text{L}}^{\prime}}^2 - M^{\text{LL}}\abs{\vec{A}_{\text{L}}}^2\right],
\label{eq:LongitudinalAction}
\end{equation}
which can be written in a form similar to \eqref{eq:TransverseAction}, provided we introduce the auxiliary field $\mathcal{A}_{\text{L}}$ as
\begin{equation}
A_{\text{L}}(\eta, \vec{k}) \equiv \frac{\mathcal{A}_{\text{L}}(\eta, \vec{k})}{f(\eta, k)}, \qquad f(\eta, k)=\frac{\sqrt{-M^{00}(\eta)}}{\sqrt{k^2 - M^{00}(\eta)}},
\end{equation}
where the corresponding element of the mass tensor reads
\begin{equation}
     M^{00}(\eta)=-a^2(\eta)\left[m^2+\gamma R(\eta)\right] + \sigma \Tilde{R}^{00}(\eta).
\label{eq:M00}
\end{equation}
Then, equation \eqref{eq:LongitudinalAction} can be expressed as
\begin{equation}
S_{\text{L}} = \frac{1}{2}\int\frac{d\eta d^3k}{(2\pi)^{3/2}}\left\{\abs{\mathcal{A}_{\text{L}}}^{\prime ~ 2} - \left[\frac{M^{\text{LL}}}{f^2} - \frac{f^{\prime\prime}}{f}\right]\abs{\mathcal{A}_{\text{L}}}^2\right\}.
\end{equation}
As before, the field can generally be written as the linear combination
\begin{equation}
\mathcal{A}_{\text{L}}(\eta, \vec{k}) = a_{\text{L}}(\vec{k})v_{\text{L}}(\eta, k) + a_{\text{L}}^*(-\vec{k})v_{\text{L}}^*(\eta, k),
\label{eq:LongitudinalExpansion}
\end{equation}
leading to the mode equation
\begin{equation}
v^{\prime\prime}_{\text{L}}(\eta, k) + \omega^2_{\text{L}}(\eta, k)v_{\text{L}}(\eta, k) = 0,
\label{eq:LongitudinalModeEquation}
\end{equation}
where the frequency is given by
\begin{equation}
\omega^2_{\text{L}}(\eta, k)=\frac{M^{\text{LL}}(\eta)}{f(\eta, k)^2} - \frac{f^{\prime\prime}(\eta, k)}{f(\eta, k)}, \qquad M^{\text{LL}}(\eta) = a^2(\eta)\left[m^2+\gamma R(\eta)\right] + \sigma \Tilde{R}^{\text{LL}}(\eta).
\label{eq:LongitudinalFrequency}
\end{equation}
We see here that, contrary to the transverse mode and the scalar field case, the shape of the longitudinal frequency is not simply $k^2$ plus an $\eta$-dependent term, but indeed more involved, and temporal and wavenumber dependencies mix. 

Note that, after canonical quantization, the coefficients of the expansions of both the transverse and the longitudinal modes will be promoted to creation and annihilation operators. As long as solutions to their respective equations of motion are normalized as $v_kv^{\prime\, *}_k - v^{\prime}_k v_k^* = i$, these operators will fulfill the standard commutation relations. For any of the spatial modes, the relation between two different basis of solutions $v_k$ and $u_k$---each of which expands the corresponding component, $A_{\text{T}}$ or $\mathcal{A}_{\text{L}}$, with different coefficients---is given in terms of the Bogoliubov transformation $u_k = \alpha_k v_k + \beta_k v_k^*$, with $\abs{\alpha_k}^2 - \abs{\beta_k}^2 = 1$, and similarly for the associated creation and annihilation operators, $\hat{b}_{\vec{k}} = \alpha_k^* \hat{a}_{\vec{k}} - \beta_k^* \hat{a}_{\vec{k}}^{\dagger}$~\cite{Birrell1982}. If the frequency is time-independent, $\beta_k=0$ and there is no particle production. However, if the geometry changes with time, two particlar solutions $v_k$ and $u_k$ will correspond, in general, to different notions of vacuum and particle \cite{Mukhanov2007}. As a consequence, the mean number density of $b$-particles in the $a$-vacuum will not be zero, but 
\begin{equation}
\bra{0^a}a_k^{\dagger, b}a_k^b\ket{0^a} = \abs{\beta_k}^2,    
\label{eq:particleproduction}
\end{equation}
from which the total mean density is obtained after integration over all wavenumbers. The latter will remain finite as long as $\abs{\beta_k}^2 \to 0$ faster than $k^{-3}$ for increasing~$k$.

One can now associate each notion of particle to observers living before and after the expansion. We consider the Bunch-Davies vacuum as the initial state, so that the observer before the expansion measures no particles. Then, the abundance of dark matter resulting from the expansion of the geometry during inflation and reheating is obtained by computing the mean number of particles an inertial observer would measure after reheating in the state of the system. If expansion is adiabatic enough after this point, particle production will be negligible. As long as the dark matter field is non-interacting, one can extrapolate this abundance to the present and compare with observations. In order to solve the mode equations~\eqref{eq:TransverseModeEquation} and \eqref{eq:LongitudinalModeEquation} to obtain the number of particles that are produced in each mode, it remains to specify the background dynamics, analyzed in the following subsection.

\subsection{Background dynamics}
\label{subsec:background}

In order to describe the geometry during inflation and reheating we consider a chaotic inflationary model consisting of a single scalar field $\phi$ whose equation of motion, assuming homogeneity and isotropy, is given by  
\begin{equation}
0 = \ddot{\phi} + 3 H(t)\dot{\phi} + \partial_{\phi}V(\phi),
\label{eq:InflatonEOMFlat}
\end{equation}
where the potential is of the form $V(\phi) = \frac{1}{2}m_{\phi}^2\phi^2$, $m_{\phi}$ denotes the inflaton mass,~$H(t) \equiv \dot{a}(t)/a(t)$ is the Hubble parameter, and the dot is derivative with respect to the cosmological time $t$.  The relation between cosmological and conformal time is obtained from $\eta = \eta_0 + \int_{0}^t d t/a(t)$. Together with \eqref{eq:InflatonEOMFlat}, we solve the corresponding Friedmann equation,
\begin{equation}
H^2 = \frac{4\pi}{3M_P^2}\left[\dot{\phi}^2 + 2V(\phi)\right],
\label{eq:HubbleRate}
\end{equation}
considering that the only contribution to the energy-momentum tensor comes from the inflaton. Therefore, the Ricci curvature scalar and the components of the Ricci tensor in terms of $\phi$ and $\eta$ read
\begin{equation}
R =\frac{8\pi}{M_P^2}\left[4V(\phi) - a^{-2}\phi^{\prime \, 2}\right], \qquad \Tilde{R}^{00}=\frac{6\pi}{M_P^2}a^{-4}\phi^{\prime \, 2}, \qquad \Tilde{R}^{ii}=\frac{2\pi}{M_P^2}a^{-4}\phi^{\prime \, 2},
\label{eq:RicciScalarTensor}
\end{equation} 
which allows us to explicitly write both \eqref{eq:TransverseFrequency} and \eqref{eq:LongitudinalFrequency}. 

\begin{figure}[t!]
\includegraphics[width=0.996\textwidth]{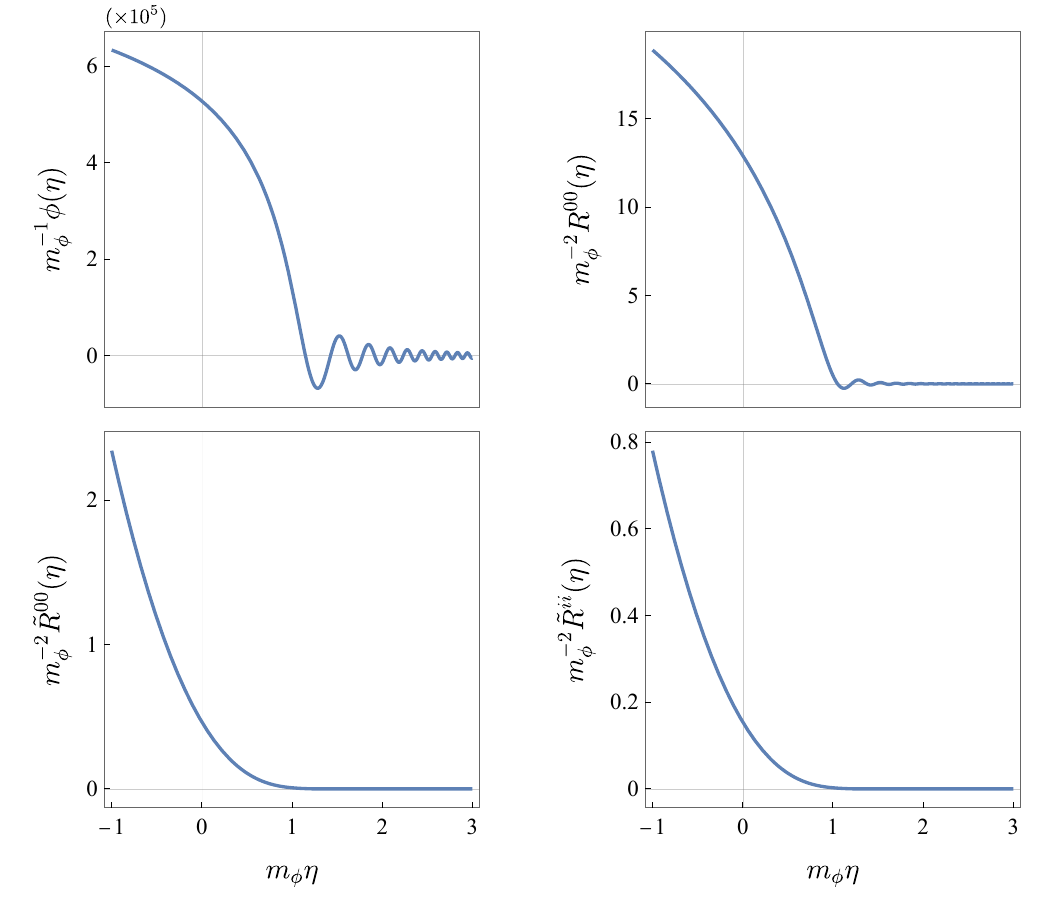}
\begin{picture}(0,0)
\put(334, 294.5){\includegraphics[width=0.205\textwidth]{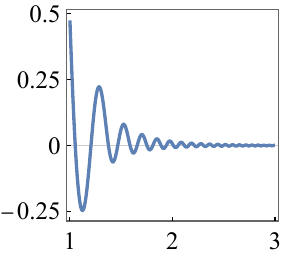}}
\end{picture}
\begin{picture}(0,0)
\put(340, 118){\includegraphics[width=0.205\textwidth]{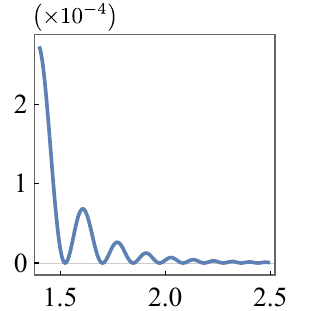}}
\end{picture}
\begin{picture}(0,0)
\put(109, 118){\includegraphics[width=0.205\textwidth]{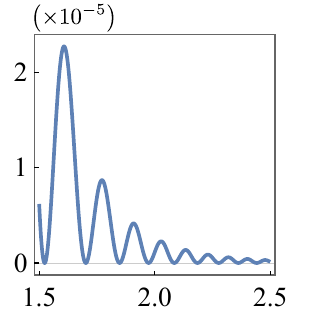}}
\end{picture}
\caption{Inflaton field $\phi(\eta)$ (upper-left panel), curvature scalar $R(\eta)$ (upper-right panel) and components of the traceless Ricci tensor (bottom row), as functions of conformal time. The range of time corresponds to the end of inflation and the beginning of reheating. The parameters used for all figures in this article are given in Appendix \ref{app:parameters}.}
\label{fig:InflatonAndRicciConformal}
\end{figure}

The dynamics is obtained by numerically solving equation \eqref{eq:InflatonEOMFlat}, together with \eqref{eq:HubbleRate}, up to the point in which particle production becomes negligible. Inflation starts at $\eta_{\text{i}}$ and ends at~$\eta=t=0$, when reheating begins, leading to the characteristic oscillations of the inflaton field~$\phi$ and, as a consequence, of the rest of the curvature-dependent quantities. This can be seen in figure \ref{fig:InflatonAndRicciConformal}. Note that, although they oscillate, the traceless Ricci tensor components have always positive sign. This fact will be important when discussing instabilities further below.

Let us now discuss in which region of parameter space our theory is well-defined, starting with the coupling $\gamma$. Stability of initial conditions (which will be analyzed further below) requires $\gamma \geq 1/6$. Any value of $\gamma$ within this region is valid, as long as $m$ and $\sigma$ are such that $M^{00}<0$. However, back-reaction has to be taken into account for very large values of~$\gamma$ (see ref. \cite{Lebedev2022} for such an analysis in the context of a non-minimally coupled scalar field), and production of particles for $\gamma \gtrsim 1$ is qualitatively similar to that of $\gamma = 1$. Therefore, we consider the range $1/6\leq \gamma\leq 1$. At the same time, we want to allow for the possibility of vanishing $\sigma$, so that we can compare with scalar field production, for example. This, together with the range of $\gamma$ discussed above, sets a limit for the smallest mass we can consider,~$m=0.5m_{\phi}$. This is clear when looking at the explicit form of $M^{00}$, given in \eqref{eq:M00}. The Ricci scalar oscillates during reheating around $0$, and therefore the mass term in $M^{00}$ has to compensate for this behavior so that $M^{00}$ remains negative. On the other hand, the traceless Ricci tensor component is always positive or $0$ (see figure \ref{fig:InflatonAndRicciConformal}). This means that a positive value of $\sigma$ is going to contribute always against the mass term. Therefore, $\sigma$ has to be negative if we want to keep the chosen range in $\gamma$ while having $m\geq 0.5m_{\phi}$. Because of all the previous considerations, we restrict ourselves to the following region of parameter space:
\begin{equation}
m \geq 0.5m_{\phi}, \qquad \gamma \in [1/6, 1], \qquad \sigma \leq 0.
\label{eq:ParameterSpace}
\end{equation}
One can consider smaller values of the mass while keeping the same allowed range in $\gamma$, but then the coupling $\sigma$ has to become negative and $\sigma=0$ is not allowed. At the same time, positive values of $\sigma$ are possible, but in this case it is required that the mass of the test field becomes greater than $0.5m_{\phi}$.

\section{Particle production of transverse modes}
\label{sec:particleproductionT}

The dynamics of the transverse modes resembles that of a scalar field, and thus we closely follow the analysis done previously in \cite{Cembranos2020, ScalarField2023}, where now we have an additional coupling to take into account.

\subsection{Solution to the transverse mode equation}
\label{subsec:transversemodeequation}

If we want to calculate \eqref{eq:particleproduction} in order to obtain the number of gravitationally produced particles in transverse modes, we need to compare at the same time the two solutions corresponding to the (Bunch-Davies) vacuum at the beginning of inflation and to the one at the time when the process is over. In particular, making the evaluation of \eqref{eq:particleproduction} at this final time requires solving equation \eqref{eq:TransverseModeEquation} from the beginning of inflation, $\eta_{\text{i}}$, until a time $\eta_{\text{f}}$ in the reheating era for which the expansion of the geometry is adiabatic enough so that there is almost no production of particles after this point. The transverse mode equation, however, has no analytical solution given the background described in subsection \ref{subsec:background}. Nevertheless, during the first stages of inflation, within the slow-roll of the inflaton field~$\phi$, one can use the analytical approximation
\begin{equation}
    v_{\text{T}, \text{SR}}(\eta, k) \simeq \sqrt{\pi\tau(\eta, k)/2} e^{i\pi\nu} H_{\nu}^{(1)}\left(k\tau(\eta, k)\right),\quad 
\tau(\eta, k)= \Bigg|\frac{\omega_{\text{T}}(\eta, k)}{\omega_{\text{T}, \text{dS}}(\eta, k)}(\eta-\eta_*) + \eta_*-\eta_0\Bigg|,
\label{eq:ApproximateTSolution}
\end{equation}
which is compatible with Bunch-Davies initial conditions and valid until $\eta_*$, whose particular value depends on the wavenumber $k$, the mass of the field $m$, and the couplings $\gamma$ and $\sigma$ (see appendix \ref{app:slowrollapproximation}) . Here, $\omega_{\text{T},\text{dS}}$ is the frequency in a de Sitter geometry, 
\begin{equation}
    \omega_{\text{T}, \text{dS}}(\eta, k)^2 = k^2 + \frac{\mu^2}{(\eta-\eta_0)^2}, \qquad \text{with} \qquad \mu^2 = m^2/H_0^2 + 12\gamma,
\end{equation}
where $H_0=H(\eta_{\text{i}})=1/\eta_0$ and $\nu = \sqrt{1/4 - \mu^2}$. This coincides, precisely, with the configuration of the geometry at the beginning of inflation. Note that if we write $\gamma=\xi-1/6$ we recover the equations for the case of gravitational production of scalar particles in the same background,~$\xi$ being the coupling to $R$. Thus, the initial vacuum state becomes unstable for $\gamma<0$ in the case of transverse mode production, being $\gamma=0$ the conformal point (equivalent to $\xi=1/6$ in the scalar field case). Equation~\eqref{eq:TransverseModeEquation} is then solved in the following two regions,
\begin{equation}
\eta = \begin{cases}
\eta_{\text{i}} \leq \eta \leq \eta_*, \quad \text{Slow-roll approximation},\\
\eta_*\leq \eta \leq \eta_{\text{f}}, \quad \text{Numerical solution},
\end{cases}
\end{equation}
where numerical computations is needed once the slow-roll approximation described above is no longer valid. Let us remark that the use of this analytical solution, studied in detail in~\cite{ScalarField2023}, is crucial for alleviating the numerical workload, especially given that the space of parameters is now three-dimensional, in contrast to the non-minimally coupled scalar case.

\begin{figure}[t!]
\includegraphics[width=\textwidth]{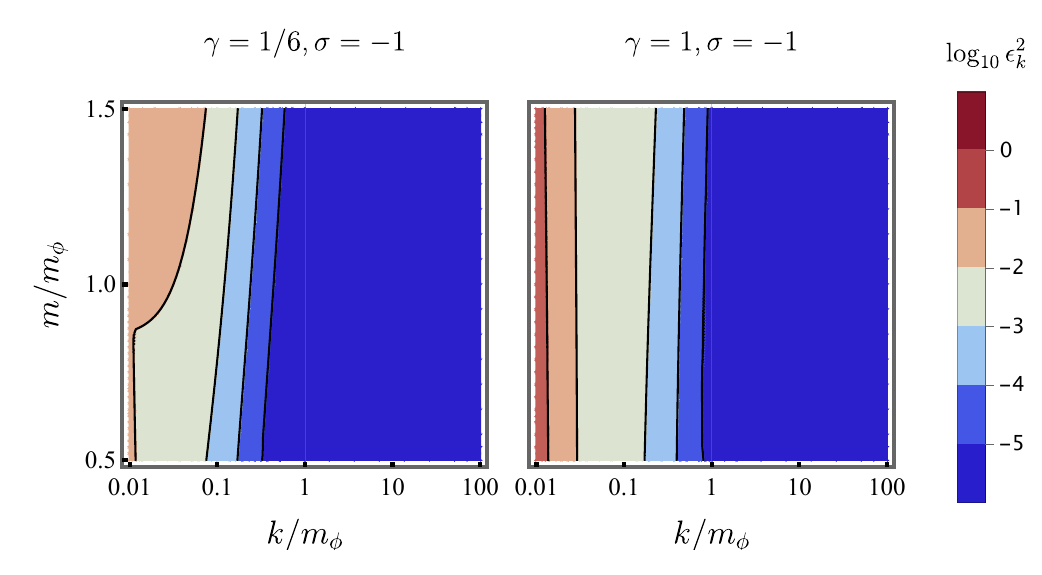}
\caption{Error squared $\epsilon_k^2$ for the transverse modes as function of the wave number $k$ and the field mass $m$, for $\gamma=1/6$ (left) and $\gamma=1$ (right), and $\sigma=-1$, for $\eta_*=-500m_{\phi}$. The behaviour for $\sigma=0$ is qualitatively similar.}
\label{fig:errorT}
\end{figure}

Then, it remains only to specify the vacuum of the observer living at $\eta_{\text{f}}$, for which we use the customary zeroth-order adiabatic prescription \cite{Birrell1982,Mukhanov2007},
\begin{equation}
\begin{split}
      u_{\text{T}}(\eta_{\text{f}}, k)&=\frac{1}{\sqrt{\omega_{\text{T}}(\eta_{\text{f}}, k)}}, \\
      u^{\prime}_{\text{T}}(\eta_{\text{f}}, k)&=-\frac{1}{\sqrt{\omega_T(\eta_{\text{f}}, k)}}\left(i\omega_{\text{T}}(\eta_{\text{f}}, k) + \frac{1}{2}\frac{\omega_{\text{T}}^{\prime}(\eta_{\text{f}}, k)}{\omega_{\text{T}}(\eta_{\text{f}}, k)}\right).
\end{split}
\end{equation}
Note that this prescription is a good notion of vacuum as long as the dynamics are adiabatic at $\eta_{\text{f}}$. The regime of adiabaticity is reached at a different point in time depending on the parameters $k, m, \gamma$ and $\sigma$, and therefore one has to choose an end point $\eta_{\text{f}}$ that fulfills the adiabaticity condition for all the parameter space considered. 

With this, we can calculate the number of produced particles, 
\begin{equation}
   n_{\text{T}}(m, \gamma, \sigma)= \int \frac{dk}{2\pi^2} k^2\abs{\beta_{\text{T}}}^2(k, m, \gamma, \sigma) = n_{\text{T,SR}} + \int_0^{\infty}\frac{dk}{2\pi^2}k^2\abs{\beta_{\text{T,SR}}}^2\mathcal{O}(\epsilon_k^2),
   \label{eq:particleproductionT}
\end{equation}
with $n_{\text{T,SR}} = \int_0^{\infty}\frac{dk}{2\pi^2}k^2\abs{\beta_{\text{T,SR}}}^2$ being the result obtained when using the approximation $v_{\text{T}, \text{SR}}$. The error in \eqref{eq:particleproductionT} due to the use of the approximate solution \eqref{eq:ApproximateTSolution} is essentially given by $\epsilon_k^2$, with $\epsilon_k$ defined in appendix \ref{app:slowrollapproximation}, and depends on the value of $k$, $m$ and $\xi$ for a fixed $\eta_*$. This is shown in figure \ref{fig:errorT} for $\eta_*=-500m_{\phi}$ on specific but representative values of the parameters for illustrative purposes. We note that $\epsilon_k^2$ is essentially independent of the value of the mass $m$ within this range, and is small for $k>m_{\phi}$. For wavenumbers smaller than the mass of the inflaton, the error increases. However, the density of produced particles is suppressed in this range by the $k^2$ factor in \eqref{eq:particleproductionT}, and therefore $n_{\text{T}}(m, \gamma, \sigma) \approx n_{\text{T,SR}}$ is a good approximation.

\subsection{Spectra of produced transverse modes}
\label{subsec:transversalspectra}

\begin{figure}[t!]
\includegraphics[width=\textwidth]{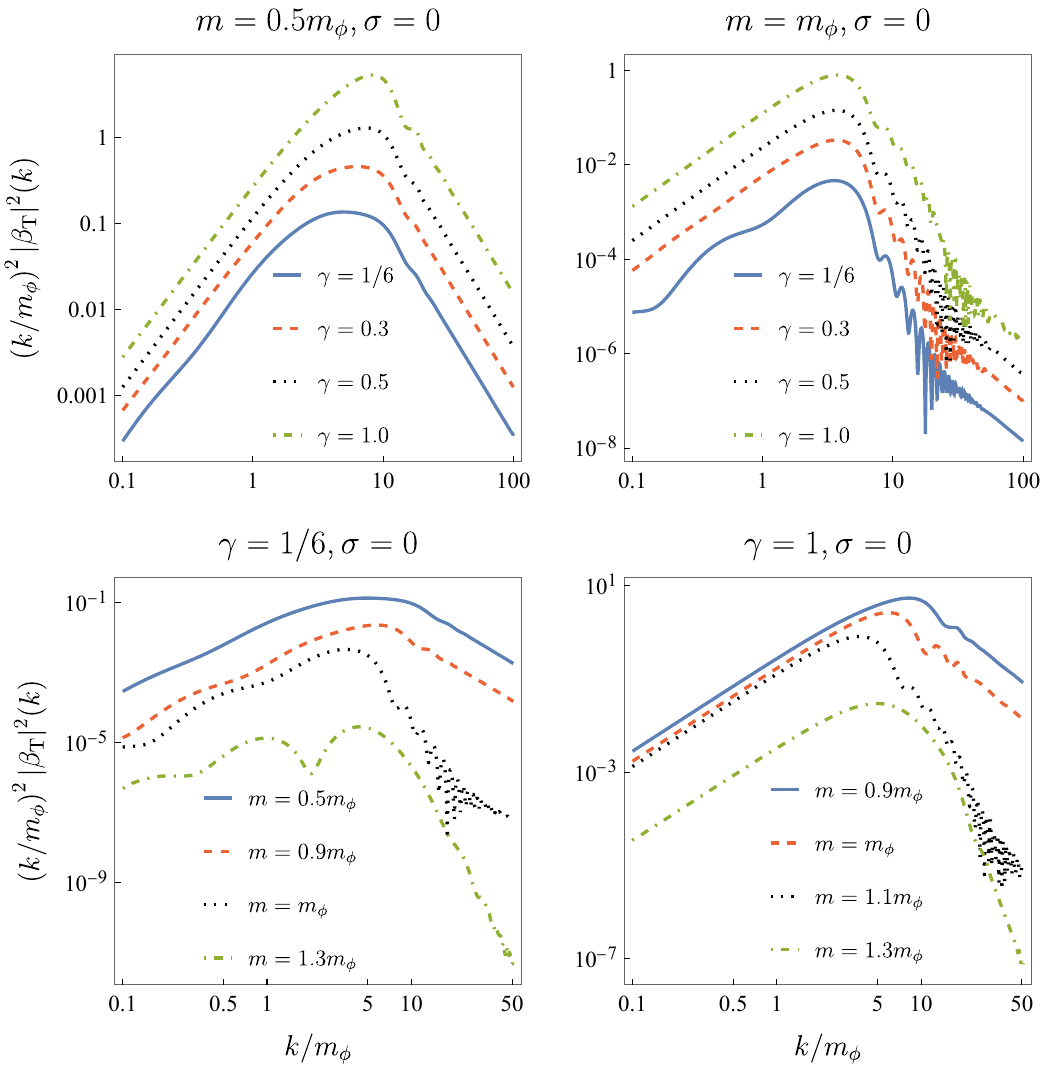}
\caption{Spectra of produced transverse modes, in log scale, for a vanishing coupling $\sigma$. The upper panels show the spectra for several values of $\gamma$, given $m=0.5m_{\phi}$ (left) or $m=m_{\phi}$ (right). On the other hand, the lower panels show the spectra for several values of $m$, given $\gamma=1/6$ (left) or $\gamma=1$ (right).}
\label{fig:SpectraT1}
\end{figure}

In this subsection we show the results of particle production of transverse modes, given the inflationary model described in subsection \ref{subsec:background}. In particular, we focus on the study of $k^2 \abs{\beta_{\text{T}}}^2$ as it is the interesting quantity for particle density production. 

We first consider a vanishing value of the coupling $\sigma$, that is to say, no coupling to the traceless Ricci tensor. The resulting spectra can be found in figure \ref{fig:SpectraT1}, where the top panels correspond to a fixed value of $m$, whereas the bottom panels are for a fixed value of $\gamma$. A logarithmic scale was used so that the UV and IR behaviour can be observed. Note that increasing the mass of the dark matter particle leads to less production in all scales. In particular, spectra for $m=m_{\phi}$ feature oscillations for large values of $k$, and fall off much faster. This falling is enhanced when the mass goes beyond this value. On the other hand, increasing the value of $\gamma$ strengthens the coupling to the geometry, and therefore particle production increases. Interestingly, the maximum of the spectra depends on both the value of~$m$ and $\gamma$.

\begin{figure}[t!]
\includegraphics[width=\textwidth]{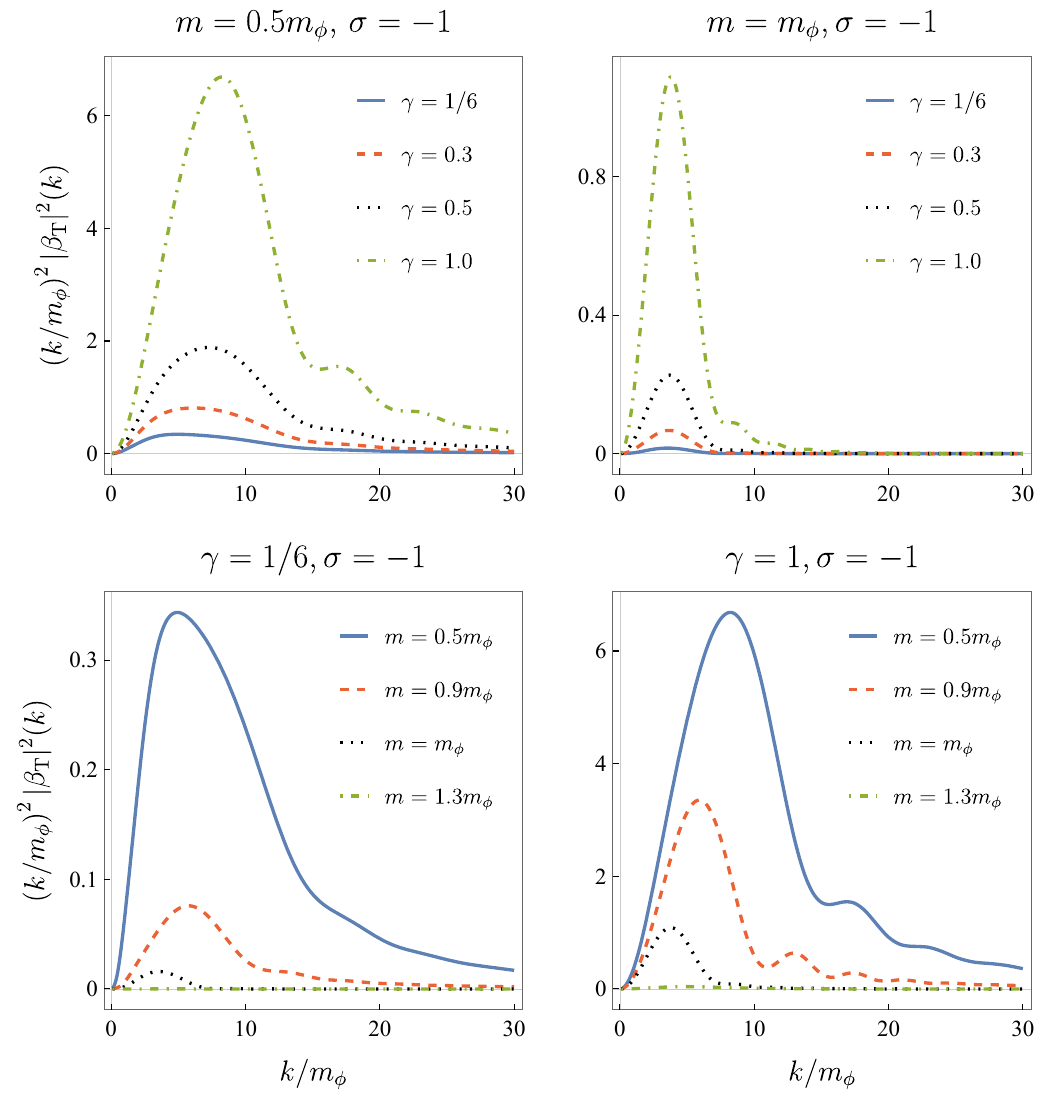}
\caption{Spectra of produced transverse modes, in linear scale, for a fixed value $\sigma=-1$. The upper panels show the spectra for several values of $\gamma$, given $m=0.5m_{\phi}$ (left) or $m=m_{\phi}$ (right). On the other hand, the lower panels show the spectra for several values of $m$, given $\gamma=1/6$ (left) or $\gamma=1$ (right).}
\label{fig:SpectraT2}
\end{figure}

Let us consider now a negative value of $\sigma$. In particular, resulting spectra for $\sigma=-1$ are depicted in figure \ref{fig:SpectraT2}. Note that a linear scale is used in this case for illustrating the magnitude of $n_{\text{T,SR}}$. The behaviour is similar to the case $\sigma=0$, but particle production is enhanced by the extra term in the frequency  \eqref{eq:TransverseFrequency}. Indeed, a negative value of $\sigma$ implies that the traceless Ricci tensor term in the frequency always contributes as a negative term (see figure \ref{fig:InflatonAndRicciConformal}). This means that it causes tachyonic instabilities as well as enhances those coming from the oscillations of the curvature scalar around $0$, which are precisely the main source of non-adiabaticity, as analyzed in \cite{Ema2016,Markkanen2017a,Ema2018,Cembranos2020, ScalarField2023}. Similarly, one also finds that a positive value of $\sigma$ yields less production, since the corresponding term contributes always against tachyonic instability. Note that this is not in contradiction with the requirement $M^{00}<0$. A large, positive value of $\sigma$ has to be compensated with the values of $m$ and $\gamma$. However, once within the validity of the theory, a negative value of $\sigma$ will contribute towards the frequency becoming imaginary.

\section{Particle production of longitudinal modes}
\label{sec:particleproductionL}

Production of longitudinal modes is more involved than that of transverse modes or a scalar field since the form of the frequency \eqref{eq:LongitudinalFrequency} is qualitatively different. However, in the limit when we approach the beginning of inflation, it can be written as a de Sitter-like frequency, which we will use to develop a slow-roll approximation similar to \eqref{eq:ApproximateTSolution}.

\subsection{Solution to the longitudinal mode equation}
\label{subsec:longitudinalmodeequation}

Let us again use the fact that the geometry approaches de Sitter at the onset of inflation, as we did in section \ref{sec:particleproductionT}. In such geometry, in which $a(\eta)=-1/H_0(\eta-\eta_0)$ and $M^{\text{LL}}=-M^{00}=\mu^2H_0^2$, the longitudinal frequency \eqref{eq:LongitudinalFrequency} becomes (we shift the conformal time $\eta-\eta_0\to\eta$ for simplicity in the notation)
\begin{equation}
    \omega_{\text{L,dS}}^2(\eta, k) = k^2 + \frac{\mu^2}{\eta^2} - \frac{k^2\left(2\eta^2k^2 - \mu^2\right)}{(\eta^2k^2+\mu^2)^2}.
\label{eq:LFrequencyDeSitter}
\end{equation}
We can now define the dimensionless quantity $x\equiv 1/(k\eta)$, so that
\begin{equation}
    \frac{\omega^2_{\text{L,dS}}(\eta, k)}{k^2} = 1 + \mu^2x^2 - \frac{x^2(2-\mu^2 x^2)}{(1+\mu^2 x^2)^2}.
\label{eq:hfunction}
\end{equation}
Note that in the limit $\mu^2x^2 \ll 1$, we can expand \eqref{eq:hfunction} in powers of $\mu^2 x^2$,
\begin{equation}
    \frac{\omega^2_{\text{L,dS}}(\eta, k)}{k^2} = 1 + \mu^2 x^2 - x^2\left(2-\mu^2 x^2\right)\left[1 - 2\mu^2x^2 + \mathcal{O}\left(\mu^4x^4\right)\right],
\end{equation}
which up to terms of order $\mathcal{O}(\mu^2x^2)$ can be written as
\begin{equation}
     \frac{\omega^2_{\text{L,dS}}(\eta, k)}{k^2} \simeq 1 + x^2(\mu^2-2).
\end{equation}
This allows us to write the frequency \eqref{eq:LFrequencyDeSitter} in this limit as
\begin{equation}
    \omega_{\text{L, dS}}^2(\eta, k) \simeq k^2 + \frac{\mu^2-2}{\eta^2},
\label{eq:LFrequencyApproximatedDeSitter}
\end{equation}
which is a good approximation for $k\eta \to k\eta_{\text{i}}$. Note that in this limit, the conformal case is, similarly to the scalar field, $\gamma=1/6$, value below which the initial vacuum becomes unstable. This is precisely the reason why we required $\gamma \geq 1/6$.

\begin{figure}[t!]
\includegraphics[width=0.996\textwidth]{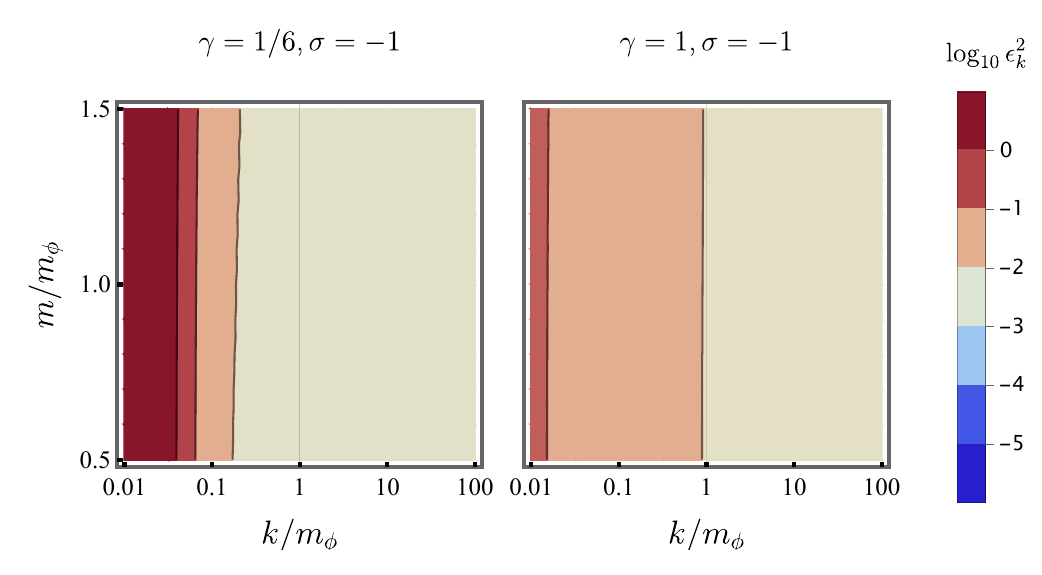}
\caption{Error squared $\epsilon_k^2$ for the longitudinal modes as function of the wave number $k$ and the field mass $m$, for $\gamma=1/6$ (left) and $\gamma=1$ (right), and $\sigma=-1$, fixed the value of $\eta_*=-500m_{\phi}$. The behaviour for $\sigma=0$ is qualitatively similar.}
\label{fig:errorL}
\end{figure}

Then, we can write an approximation to the solution during slow-roll for the longitudinal mode equation as
\begin{equation}
    v_{\text{L}, \text{SR}}(\eta, k) \simeq \sqrt{\pi\tau(\eta, k)/2} e^{i\pi\nu} H_{\nu}^{(1)}\left(k\tau(\eta, k)\right),\quad 
\tau(\eta, k)= \Bigg| \frac{\omega_{\text{L}}(\eta, k)}{\omega_{\text{L}, \text{dS}}(\eta, k)}(\eta-\eta_*) + \eta_*-\eta_0\Bigg |,
\label{eq:ApproximateLSolution}
\end{equation}
where the de Sitter frequency is now given by (reintroducing $\eta_0$)
\begin{equation}
    \omega_{\text{L}, \text{dS}}(\eta, k)^2 = k^2 + \frac{\mu^2-2}{(\eta-\eta_0)^2}, \qquad \text{with} \qquad \mu^2 = m^2/H_0^2 + 12\gamma.
\end{equation}
From this point on, the analysis follows closely what has been discussed in section \ref{sec:particleproductionT}. In particular, the error in the number density of produced longitudinal modes due to the use of~\eqref{eq:ApproximateLSolution} is shown in figure~\ref{fig:errorL} for the same value $\eta_*=-500m_{\phi}$. Note that for these modes, the slow-roll approximation is slightly worse, although the behaviour is qualitatively similar to that of the transverse modes. However, the error becomes important again for low values of $k$, which are suppressed in the density, and the integral of the spectra is not significantly affected. 

\subsection{Spectra of produced longitudinal modes}
\label{subsec:longitudinalspectra}

Let us now show the particle production spectra for longitudinal modes, according to the background dynamics discussed in subsection \ref{subsec:background}. As before, we analyze the quantity $k^2 \abs{\beta_{\text{T}}}^2$ for different values of the parameters. 

\begin{figure}[t!]
\includegraphics[width=0.9\textwidth]{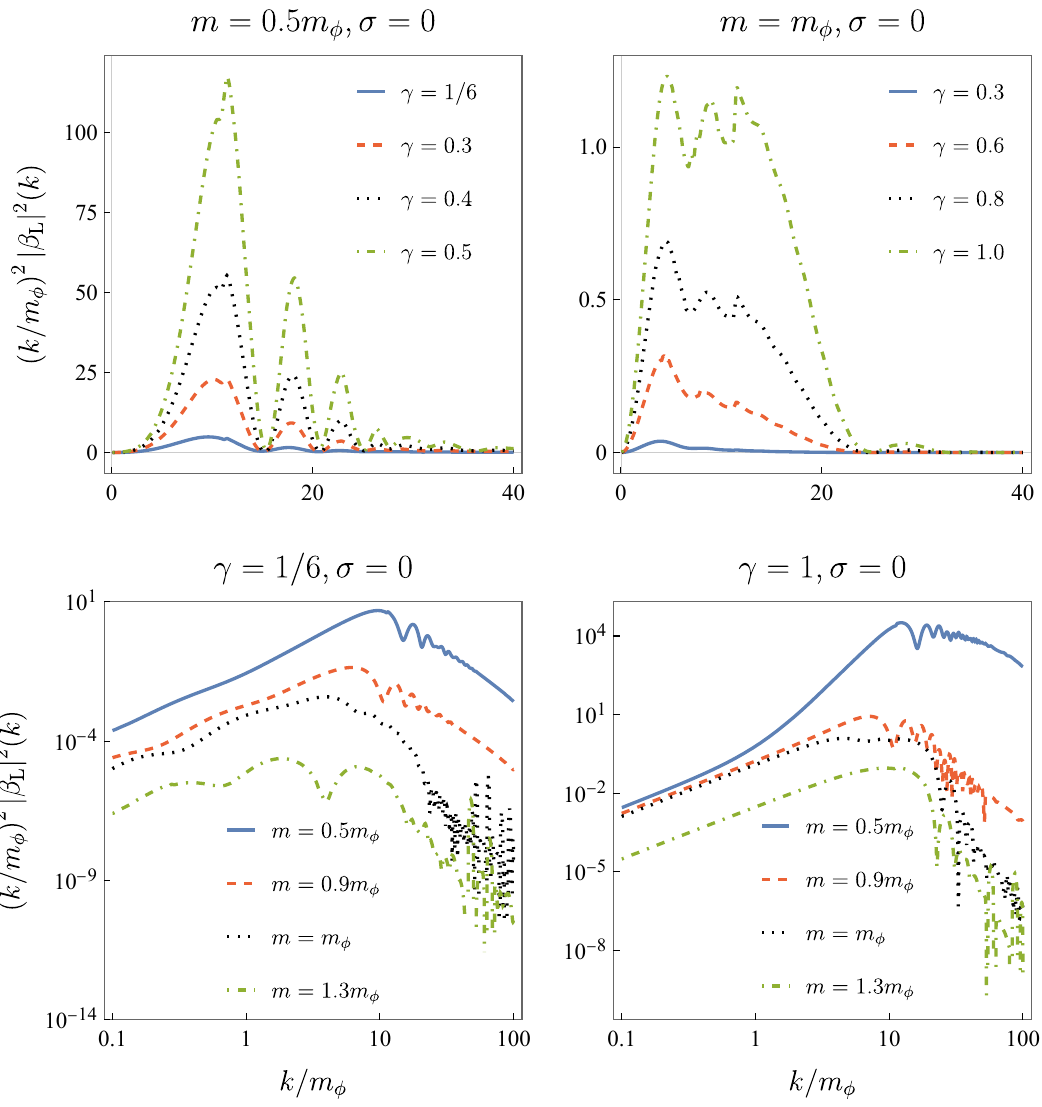}
\caption{Spectra of produced longitudinal modes for a vanishing coupling $\sigma$. The upper panels, in linear scale, show the spectra for several values of $\gamma$, given $m=0.5m_{\phi}$ (left) or $m=m_{\phi}$ (right). On the other hand, the lower panels, in log scale, show the spectra for several values of $m$, given $\gamma=1/6$ (left) or $\gamma=1$ (right).}
\label{fig:SpectraL1}
\end{figure}

Because the shape of the frequency is different, the spectra of produced particles are qualitatively different from those obtained in the case of transverse modes. Let us analyze the same cases in the same order.
We first concentrate in the situation in which the $\sigma$ coupling vanishes, represented in figure \ref{fig:SpectraL1}. Contrary to the transverse case, all spectra feature significant oscillations in $k$, not only for $m=m_{\phi}$. As before, for fixed mass (top panels), particle production increases with the coupling $\gamma$, whereas it increases when decreasing the dark matter mass (bottom panels) given a fixed value of $\gamma$. Crucially, spectra for $m=m_{\phi}$ fall off faster, as it was observed before. Note that, in general, production of particles is much more important in the longitudinal mode than in the transverse ones: Spectra are broader and maxima can be several orders of magnitude larger. These effects are particularly noticeable for small masses and large values of the coupling $\gamma$, i.e., in scenarios in which production is enhanced (see top left and bottom right panels of figure \ref{fig:SpectraL1}).

\begin{figure}[t!]
\includegraphics[width=\textwidth]{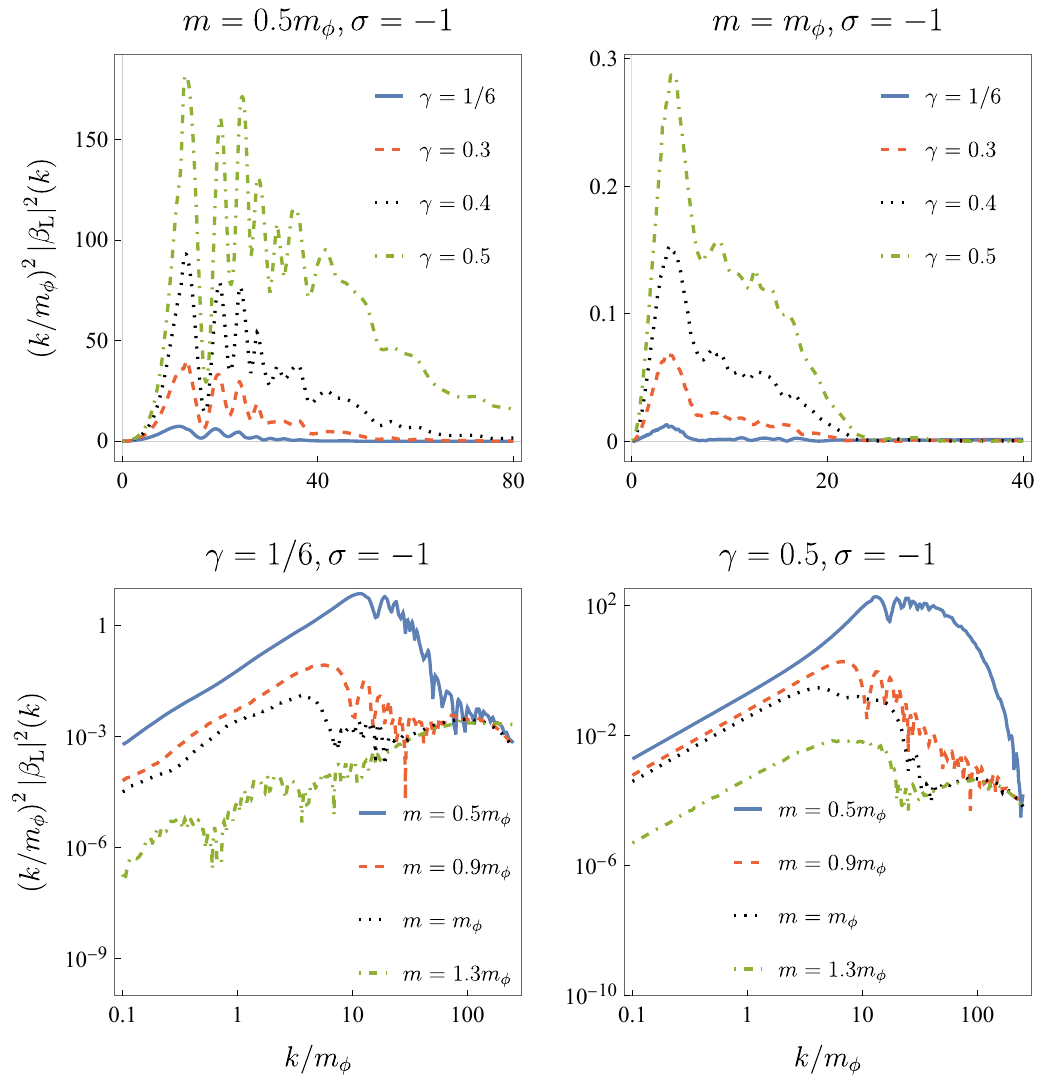}
\caption{Spectra of produced longitudinal modes for a fixed value $\sigma=-1$. The upper panels, in linear scale, show the spectra for several values of $\gamma$, given $m=0.5m_{\phi}$ (left) or $m=m_{\phi}$ (right). On the other hand, the lower panels, in log scale, show the spectra for several values of $m$, given $\gamma=1/6$ (left) or $\gamma=0.5$ (right).}
\label{fig:SpectraL2}
\end{figure}

As anticipated, non-vanishing, negative values of $\sigma$ will induce tachyonic instabilities, yielding larger production. This is illustrated in figure \ref{fig:SpectraL2}, where one can observe that oscillations in the spectra are much more violent as well, as compared to the $\sigma=0$ case. Because spectra are wider, we resolve them up to $k\approx250m_{\phi}$ and restrict ourselves to $\gamma\leq 0.5$ in order to calculate the total density of particles produced. Similarly to the transverse modes, a positive value of $\sigma$ would lead to less production, since instabilities would become less important. This amplification of the tachyonic behavior of the frequency through the coupling $\sigma$ is absent in the non-minimally coupled scalar field case for obvious reasons, but it is an important mechanism for producing a large abundance of dark matter. We therefore devote the next section to the analysis of this quantity.

\section{Abundance of produced particles}
\label{sec:abundances}

\begin{figure}[t!]
\includegraphics[width=0.99\textwidth]{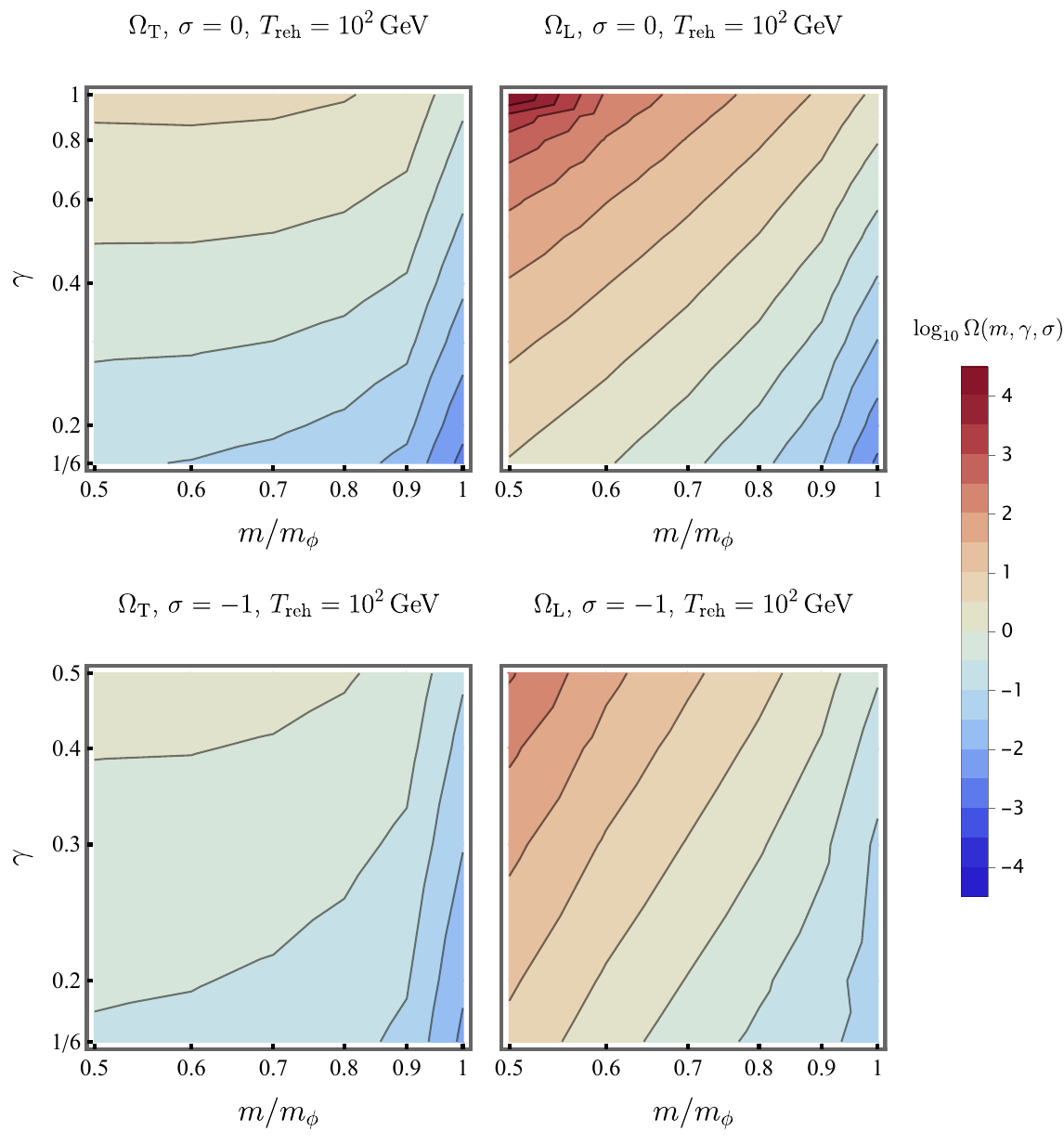}
\caption{Transverse (left panels) and longitudinal (right panels) abundances for $\sigma=0$ (upper panels) and $\sigma=-1$ (bottom panels), and a reheating temperatures of $T_{\text{reh}}=10^2 \, \text{GeV}$. Note that $\Omega_{\text{T}}$ includes only one transverse degree of freedom.}
\label{fig:IndividualAbundances}
\end{figure}

\begin{figure}[t!]
\includegraphics[width=0.99\textwidth]{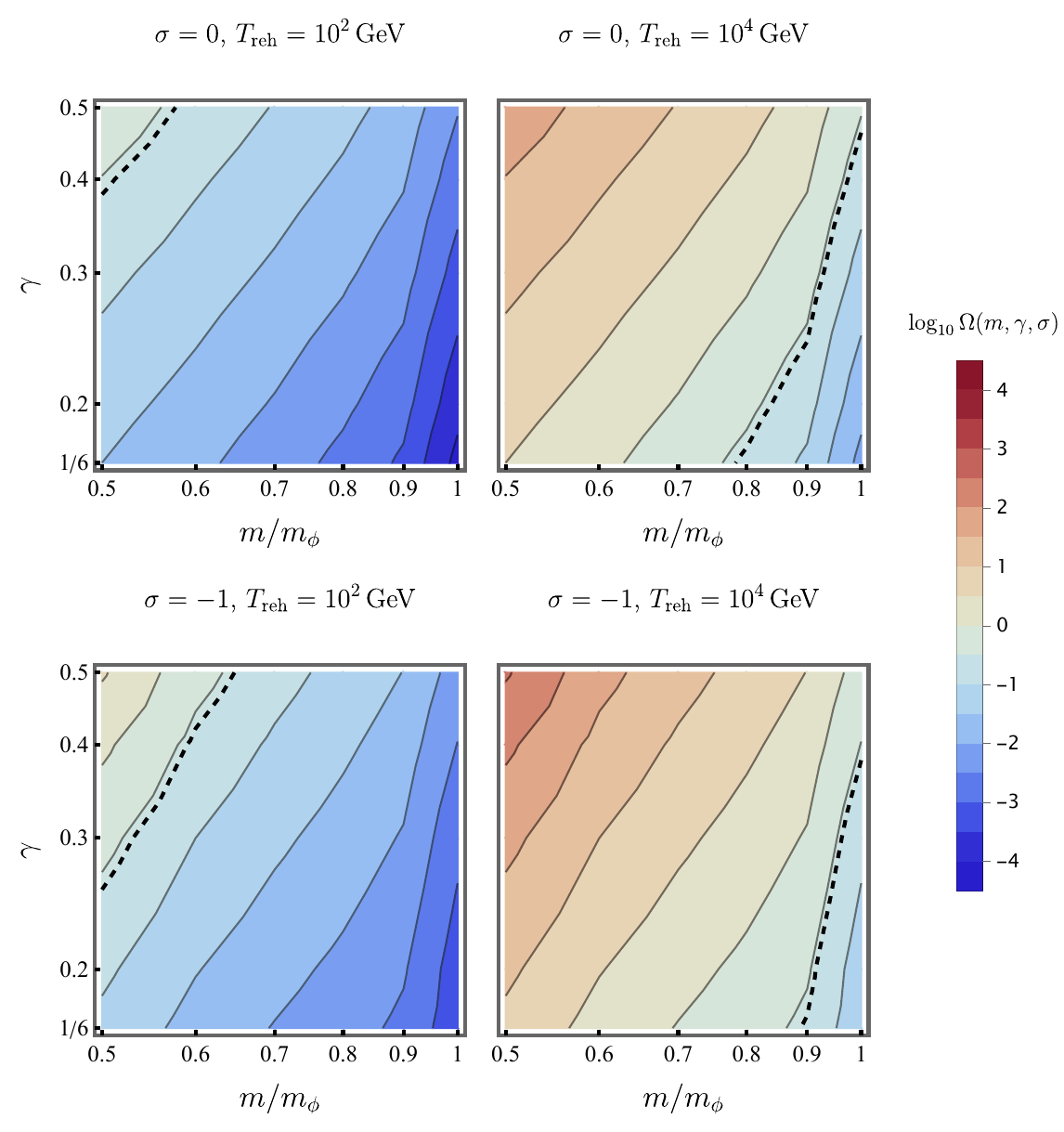}
\caption{Total abundance for $\sigma=0$ (upper panels) and $\sigma=-1$ (bottom panels), and two different reheating temperatures, $T_{\text{reh}}=10^2 \, \text{GeV}$ (left panels) and $T_{\text{reh}}=10^4 \, \text{GeV}$ (right panels). The dashed line corresponds to the observed dark matter abundance.}
\label{fig:TotalAbundances}
\end{figure}

As we have seen, the mechanism of gravitational particle production discussed above results in a non-vanishing density of particles of the test field $A_{\mu}$. The total particle density is obtained after integrating $k^2\abs{\beta_k}^2$ over all values of $k$, as described in equation \eqref{eq:particleproductionT}. If the dark matter field is non-interacting (i.e., it is a spectator field), and particle production becomes negligible after $\eta_{\text{f}}$, this comoving density will remain the same until today. In other words, it will be related to the physical density only by a factor that takes into account the dilution due to the expansion of the universe, $n_{\text{T,phys}}(\eta_{\text{today}})a^3(\eta_{\text{today}})=n_{\text{T}}$ (and similarly for the longitudinal mode). Writing the abundance today in terms of the background radiation temperature, one arrives at \cite{Cembranos2020,ScalarField2023}
\begin{equation}
    \Omega (m, \gamma, \sigma) = \frac{8\pi}{3M_P^2H^2_{\text{today}}}\frac{g_{*S}^{\text{today}}}{g_{*S}^{\text{reh}}}\left(\frac{T_{\text{today}}}{T_{\text{reh}}}\right)^3 m \, \frac{n(m, \gamma, \sigma)}{a_{\text{reh}}^3},
\end{equation}
where $T_{\text{today}}$, $T_{\text{reh}}$ and $g^{\text{today}}_{*S}$, $g^{\text{reh}}_{*S}$ are the radiation temperatures today and at the end of reheating, and the corresponding relativistic degrees of freedom, respectively.
Note that~$n = n_{\text{L}}+2n_{\text{T}}$, in order to include in the abundance the production for the three modes of the vector field.
Additionally, the value of the scale factor at the end of reheating can be obtained as function of the reheating temperature, which is a free parameter. Indeed, when radiation dominates, at~$\eta_{\text{reh}}$, the Hubble rate can be expressed as
\begin{equation}
    H^2_{\text{reh}} = \frac{8\pi}{3M_P^2}\frac{\pi^2}{30}g_{*S}^{\text{reh}}T_{\text{reh}}^4,
\end{equation}
thus allowing to write $\eta_{\text{reh}}$ and $a_{\text{reh}}$ as function of $T_{\text{reh}}$. Let us remark that in the whole parameter space considered, adiabaticity is reached way before the end of reheating (namely~$\eta_{\text{f}} < \eta_{\text{reh}}$) as long as $T_{\text{reh}} \lesssim 10^{13} \, \text{GeV}$, thus providing an upper limit for the reheating temperature.

We show the longitudinal and transverse abundances in figure \ref{fig:IndividualAbundances}, for two values of the coupling, $\sigma=0$ and $\sigma=-1$, and fixed reheating temperature $T_{\text{reh}}=10^2 \, \text{GeV}$. As expected from the spectra analysis, the abundance of longitudinal modes is much larger, several orders of magnitude even, depending on the values of $m$ and $\gamma$. At the same time, a negative value of the coupling $\sigma$ increases both abundances by around one order of magnitude. It is also interesting to note that the transverse abundance changes slower with mass, until $m \sim m_{\phi}$ is reached. More importantly, there is no qualitative change in the behavior of the longitudinal mode abundance in the region close to the conformal point $\gamma=1/6, \sigma=0$, in contrast to the scalar field studied in \cite{ScalarField2023}.

On the other hand, total abundances are depicted in figure \ref{fig:TotalAbundances}, for $\sigma=0$ (upper panels) and~$\sigma=-1$ (bottom panels) as well as for two different reheating temperatures, $T_{\text{reh}}=10^2 \, \text{GeV}$ and $T_{\text{reh}}=10^4 \, \text{GeV}$. As before, a negative value $\sigma$ increases the total abundance (which is dominated by the longitudinal contribution), and so does increasing the value of the reheating temperature. We observe that gravitational production is able to reproduce observations for masses below the inflaton, and enhancing production (for example via a negative coupling $\sigma$) shifts the observed abundance towards larger masses. In general, gravitational production of vector fields is much more efficient than that of scalar fields, since one is able to explain observations for a dark matter candidate with a mass of the order of the inflaton mass, in contrast to the case analyzed in \cite{ScalarField2023}. In particular, note that the reheating temperatures in figures \ref{fig:IndividualAbundances} and \ref{fig:TotalAbundances} are much lower than standard values. If we were to reproduce the observed abundance for a typical value of $T_{\text{reh}}$ (for instance, $T_{\text{reh}} \sim 10^9 \, \text{GeV}$), the mass of the dark matter candidate would have to be orders of magnitude higher than the inflaton mass.

\section{Conclusions}
\label{sec:conclusions}

Gravitational production during inflation and reheating can be used as a mechanism to constrain the properties of spectator dark matter by comparing the corresponding abundance with observations. Previous works have concentrated mostly on scalar spectator fields, although minimally coupled massive vector fields have also been considered.

In this work, we have studied particle production of a massive vector boson that is coupled to both the Ricci scalar and the Ricci tensor, in the context of inflation and reheating, extending therefore previous works on the subject. We showed that such interactions can be rewritten in terms of a mass tensor that results in an effective, time-dependent mass in the equations of motion of the respective modes. Crucially, the fact that this effective mass has to be always possitive restricts the valid parameter space of the theory. In this way, the dynamics of a Proca field are generalized to incorporate the evolution of the relevant background. The latter is given by numerically solving the equation of motion of the inflaton field, without the need of assuming slow-roll, for a particular inflationary potential, although the procedure is completely independent of this choice. We made use of an analytical approximation to the solution of the mode equations of both transverse and longitudinal modes, valid during the slow-roll part of inflation, which allowed us to increase efficiency of calculations and to properly define initial conditions for the spectator field. As a result, we presented the spectra of produced particles together with the corresponding abundances obtained after dilution of the total density of produced particles after the adiabatic regime has been reached. The longitudinal mode production dominates over that of the transverse modes, being the leading contribution to the final abundance. As expected, production increases when decreasing the test field mass, and increasing the coupling to the curvature scalar, following a pattern similar to scalar fields. However, the coupling to the Ricci tensor can take negative values, leading to tachyonic instabilities that facilitate highly efficient particle production. This feature---absent when studying gravitational production of non-minimally coupled scalar fields---is especially relevant for the longitudinal case, which is generally more sensitive to instabilities.

Overall, gravitational production of such a massive vector field is much more efficient and much more influenced by tachyonic behavior than that of scalar fields. As a consequence, the observed abundance is recovered for a heavy dark matter candidate (close to the inflaton mass) only if reheating temperatures are lower than typical values. Indeed, if one considers usual reheating temperatures, the mass of the dark matter particle would have to be orders of magnitude higher than that of the inflaton, since lighter particles would be overproduced.

\section*{Acknowledgements}

This work was partially supported by the MICINN (Ministerio de Ciencia e Innovación, Spain) projects  PID2019-107394GB-I00/AEI/10.13039/501100011033 (AEI/FEDER, UE), PID2020-118159GBC44, and PID2022-139841NB-I00, COST (European Cooperation in Science and Technology) Actions CA21106 and CA21136. Additionally, Á. P.-L. is supported by the MIU (Ministerio de Universidades, Spain) fellowship FPU20/05603. J. M. S. V. acknowledges the support of the Spanish Agencia Estatal de Investigaci\'on through the grant “IFT Centro de Excelencia Severo Ochoa CEX2020-001007-S". Finally, J. A. R. C. acknowledges support by Institut Pascal at Université Paris-Saclay during the Paris-Saclay Astroparticle Symposium 2022, with the support of the P2IO Laboratory of Excellence (program “Investissements d’avenir” ANR-11-IDEX-0003-01 Paris-Saclay and ANR-10-LABX-0038), the P2I axis of the Graduate School of Physics of Université Paris-Saclay, as well as IJCLab, CEA, APPEC, IAS, OSUPS, and the IN2P3 master project UCMN.

\appendix

\section{Parameters}
\label{app:parameters}

In this appendix, we provide all the parameters which were used in the numerical calculations. Most of the results are left in terms of the mass of the inflaton, $m_{\phi}$. When necessary, we have taken $m_{\phi} = 1.2 \times 10^{13} \, \text{GeV}$, value that sets the scale of the problem. Therefore, the Planck mass $M_P$ has the value $M_P = 1.02 \times 10^6 m_{\phi}$. 

The inflaton field has an initial value of $\phi_i = 3M_P$. Inflation ends at $t = 0$, point at which, under the slow-roll approximation, the field woudl reach the value $\phi_0 = 0.5M_P$. Therefore, one can extract the initial time of inflation, $t_i \simeq -15.35/m_{\phi}$. For solving the exact equation of motion \eqref{eq:InflatonEOMFlat}, we take $\phi^{\prime}(t_i)=\phi_{\text{SR}}^{\prime}(t_i)$, i.e., the value of the derivative as given by the slow-roll approximation with the conditions imposed above. Thus, $\phi(t=0)$ slightly deviates from $\phi_0$. Furthermore, we make the choice $a(t=0)=a_0=1$. Lastly, we assume slow-roll is a good approximation until $\eta_*=-500/m_{\phi}$.

Particle production is obtained at $\eta_{\text{f}} = 7.58/m_{\phi}$, which is within the adibatic regime for all the parameter space explored, namely $0.5m_{\phi}\leq m$, $1/6\leq\gamma\leq 1$ and $\sigma \leq 0$.

\section{Slow-roll approximation for the solution to the mode equation}
\label{app:slowrollapproximation}

Let us elaborate on the approximation for the solution to the mode equations \eqref{eq:ApproximateTSolution} and \eqref{eq:ApproximateLSolution}. In what follows, we write $\eta-\eta_0$  as $\eta$, and drop the mode index $k$ for clarity. The derivation, detailed in [Out Ref.], requires that the following quantities are small, given values of $k, m, \gamma$ and $\sigma$. 
\begin{itemize}
\item Let us first define
\begin{equation}
    \epsilon(m, \gamma, \sigma) = \underset{ \eta\in I_1}{\text{max}}\Bigg|1-\frac{\omega_{{\text{SR}}}(  \eta; m, \gamma, \sigma)}{\omega_{{\text{dS}}}( \eta; m, \gamma, \sigma)}\Bigg|, \quad \text{with} \quad I_1 = (-\infty, \eta_1),
\end{equation}
where $\eta_1$ is chosen such that $\epsilon \ll 1$. Then, we can define $h(\eta; m, \gamma, \sigma)$ by 
\begin{equation}
    \frac{\omega_{\text{SR}}}{\omega_{\text{dS}}} = 1 + \epsilon h.
\end{equation}
By construction, $\abs{h(\eta)} \leq 1$ for $\eta \in I_1$. Moreover, $h^{\prime}(\eta) \geq 0$.

\item Second, we define the quantity
\begin{equation}
    \delta(m, \gamma, \sigma) = \underset{\eta\in I_2}{\text{max}}\Big|h^{\prime}(\eta; m, \gamma, \sigma)\eta\Big|, \quad \text{with} \quad I_2 = (-\infty, \eta_2),
\end{equation}
where we choose $\eta_2$ such that $\delta \leq \epsilon$. Then, we introduce $g(\eta; m, \gamma, \sigma)$ as
\begin{equation}
    h^{\prime}(\eta) = \frac{\delta g(\eta)}{\eta},
\end{equation}
for which again we have that $\abs{g_k(\eta)}\leq 1$ for $\eta \in I_2$.

\item Finally,
\begin{equation}
    \rho(m, \gamma, \sigma) = \underset{\eta \in I_3}{\text{max}} \Bigg|\frac{\omega^{\prime}_{\text{dS}}(\eta)}{\omega_{\text{dS}}(\eta)}\eta \Bigg|, \quad \text{with} \quad I_3 = (-\infty, \eta_3),
\end{equation}
and choose $\eta_3$ such that $\rho \leq \epsilon$.

Now, we take $\eta_*=\text{min}(\eta_1, \eta_2, \eta_3)$ and $I=(-\infty, \eta_*)$, where $I$ is the interval for which the three parameters $\epsilon, \delta, \rho$ are small. Note that $\eta_*<0$ since inflation ends at $\eta=0$. 

\item Additionally, we require that $|\eta_*/\eta_0|>1$.
\end{itemize}

Provided the above is fulfilled, and translation of $\eta$ to $\eta-\eta_0$ is undone, one can write the solution to the mode equation as \eqref{eq:ApproximateTSolution} and \eqref{eq:ApproximateLSolution} up to terms of order $\mathcal{O}(\epsilon^2)$, where by $\epsilon$ we mean here the largest of the three small parameters defined above.


\bibliographystyle{JHEP.bst}
\bibliography{references.bib}

\end{document}